# Oscillations in a neurite growth model with extracellular feedback

V.I. Mironov[1*], A.S. Romanov[1], A.Yu. Simonov[1], M.V. Vedunova[1] and V.B. Kazantsev[1,2]

[1]Nizhny Novgorod Neuroscience Centre, Lobachevsky State University of Nizhny Novgorod, Nizhny Novgorod, Russia
[2]Laboratory of Nonlinear Dynamics of Living Systems, Institute of Applied Physics of Russian Academy of Science, Nizhny Novgorod, Russia

Correspondence:
Mr. Vasily Mironov
Nizhny Novgorod Neuroscience Centre
Lobachevsky State University of Nizhny Novgorod
Gagarin ave.23
Nizhny Novgorod, NN, 603950, Russia
mironov@neuro.nnov.ru

**Short title**:
Neurite dynamics and oscillation model

**Abstract**

We take into account the influence of extracellular signalling on neurite elongation in a model of neurite growth mediated by building proteins (e.g. tubulin). Tubulin production dynamics was supplied by a function describing the influence of the extracellular signalling that can promote or depress the elongation. We found that such extracellular feedback can generate neurite length oscillations with periodic sequence of elongations and retractions. The oscillations prevent further outgrowth of the neurite which becomes trapped in the non-uniform extracellular field. We analyzed the characteristics of the elongation process for different distributions of attracting and repelling sources of the extracellular signal molecules. The model predicts three different scenarios of the neurite development in the extracellular field including monotonic and oscillatory outgrowth, localized limit cycle oscillations and complete depression of the growth.

**Keywords: neurites, tubulin, growth factors, cell morphology, oscillatory dynamics**

## 1. Introduction

Neural development and dendritic morphogenesis underlie formation of specific network structures, synaptic connectivity and information processing in the brain[1–4]. Abnormalities in neuronal development and regeneration are implicated in several neurological disorders such as autism, schizophrenia and epilepsy[5–9].Generation of certain morphological patterns involves complex intracellular molecular cascades modulated by extracellular signaling. Neurite elongation and branching provide the formation of certain dendritic patterns guided by extracellular growth factor molecules released by the other cells. The most significant neurotrophic factors that stimulate neurite outgrowth are GDNF, BDNF and NGF[10–12]. An inverse process called retraction is also important fordevelopment and functioning of the nervous system. It may be caused by lysophosphatidic acid [13], some types of signaling molecules such as semaphorins, netrins, and ephrins[14,15], glutamate[16] and other[17]. The elongation process depends on building proteins (e.g. tubulin, actin) produced in cell soma. Those proteins are transported to the growth cone by diffusion and active transport, then assembled in microtubules providing elongation of the neurite. In space the neurite is guided by growth factors providing the direction of the growth. The growth can be also influenced by many factors including cell adhesion, binding to extracellular matrix components[18,19].

Many mathematical models were proposed to simulate the neural development features (reviewed in [4]). Based on microtubule assembly and depolymerization the models described the processes of production, degradation, and transport of building proteins. The neurite length dynamics was based on either evolution of tubulin concentration in the neurite compartments including the growth cone [20–22], or the evolution of the tubulin spatial concentration profile along the neurite [23,24], or focused on description of mechanical properties of the neurite outgrowth [19,25,26]. Model of neurite outgrowth proposed in [27] was based on assumption that the elongation can be controlled by membrane expansion and endocytosis. Depending on coupling between the microtubules and the vesicle dynamics, different regimes corresponding to dendritic and axonal growth were found. The models mentioned above merely incorporated intracellular processes underlying the outgrowth. The extracellular signaling described in other modeling studies accounted for different aspects of neural morphogenesis such as neurite branching and navigation (e.g. [28]). However, the extracellular influence on the elongation/retraction processes was poorly understood.

In this Letter we propose the neurite growth model capable of neurite elongation and retraction and driven by extracellular signaling. Such signals coming from extracellular space may activate specific pathway regulating different aspects of neuronal development (axon and dendrite growth), synapse formation and plasticity[29]. To model the intracellular dynamics we used the tubulin-based compartmental model [22]. The extracellular signals are accounted as a non-uniform field of molecules released by neighboring cells and sensed by the growth cone. We assumed that these molecules generated a feedback signal changing the rate of tubulin production. Such changes depend on the type of extracellular molecules (attracting or repelling the growth) and, hence, promote or depress the growth. We analyzed computational consequences of such extracellular feedback and found that it can significantly affect the growth dynamics. In particular, certain profile of the extracellular concentration may induce spontaneous the neurite length oscillations. Such oscillations may delay the neurite growth or the neurite can be completely trapped oscillating with a certain frequency in a certain space region.

## 2. Materials and Methods

We investigate tubulin-based compartmental model proposed in [22] and supplied with extracellular feedback as follows:

$$\begin{cases} \dfrac{dC_0}{dt} = I - \gamma_0 C_0 + \hat{D}_{01}(C_1 - C_0); \\ \dfrac{dC_i}{dt} = \hat{D}_{i,i-1}(C_{i-1} - C_i) + \hat{D}_{i,i+1}(C_{i+1} - C_i); \\ \dfrac{dC_n}{dt} = -\gamma_n C_n + \hat{D}_{n,n-1}(C_{n-1} - C_n) - \alpha \cdot C_n + \beta; \\ \dfrac{dL}{dt} = \alpha \cdot C_n - \beta; \\ \dfrac{dI}{dt} = -\delta(I - I^*) + F(S(x,y,t)). \end{cases} \tag{1}$$

Here, variables $C_i$ describe the concentration of available (free) tubulin in the $i$th segment of the neurite ($i$=0,1,…,$n$), $L$ is the variable length of the neurite. Parameter $\gamma$ determine tubulin degradation rate, $\alpha$ is an association constant, $\beta$ is a dissociation constant, $\hat{D}_{ij} = \dfrac{DA_{ij}}{V_i \Delta x_{ij}}$ is the diffusion rate from segment $j$ to segment $i$, $A_{ij}$ is the cross section area between segments, $V_i$ is the volume of segment *i*, $\Delta x_{ij}$ is distance between the centers of the nearest-neighbor segments, *D* is the diffusion constant.

Neurotrophic factors affect the cell metabolism via growth cone. After binding to the receptor they form a complex which activates the PI3-kinase and Akt. Akt activation leads to CREB-mediated synthesis of actin and tubulin accelerating outgrowth. In the growth cone PI3-kinase also induces Ras, which is a pathway stimulating actin polymerization [11,12,17].In our model the extracellular feedback is provided by the last equation. We assume that the rate of tubulin production, *I*, is variable relative to an equilibrium level, $I^*$, with characteristic time scale, $1/\delta$. The influence of extracellular signaling is described by the function $F(S(x,y,t))$, where $S(x,y,t)$is an effective (average) concentration of the growth factors at the growth cone in the point $(x,y)$ of the *2D* area at time moment *t*. We suppose that sources of extracellular molecules (e.g. other cells) are distributed in the *2D* area and characterized by some parameter *p*. They can either promote the growth of the neurite ($p=+1$) or depress it ($p=-1$). The concentration of signaling molecules diffused from each source to the growth cone can be calculated by the following formula [30]:

$$C^f(r,t)=\frac{q}{4\pi D_1 r}\cdot\left(1-\frac{2}{\sqrt{\pi}}\int_0^{\frac{r}{\sqrt{4D_1 t}}} e^{-x^2}dx\right), \tag{2}$$

where *q* describes the constant rate of the growth factor production, *r* is distance between the source and the growth cone, and $D_1$ is the extracellular signaling molecules diffusion rate. Next, we calculated the overall influence of the molecules coming from all sources as a sum weighted by the signs, $p_k$, of corresponding sources:

$$S(x,y,t)=\sum_{k=1}^{N} p_k\cdot C_k^f(r(x,y),t). \tag{3}$$

We estimate the action of this effective concentration on the growth cone by activation function $F(S)$ taken here, for illustration, in the form of logistical curve:

$$F(S)=A\left(\frac{1}{1+\exp\left(\frac{-2\pi(S-B)}{C}\right)}-\frac{1}{2}\right), \tag{4}$$

where *A* is the maximal rate of the extracellular feedback, *B* is the midpoint of the feedback activation, *C* is the width of the active concentration interval.

The model (1)-(4) was initiated with a definite distribution of the growth factor sources. Simulation of neurite growth (1) started simultaneously with the diffusion of signaling molecules (2)-(4). Then, the concentration $S(x,y,t)$ was calculated in the reference frame moving with growth cone.

In the absence of the extracellular feedback, e.g. for $I=I^*=const$ the elongation dynamics is quite simple and was explained in details in [22]. The neurite length monotonically grows with the concentration in terminal segment decaying to a certain value defined by the tubulin production rate (Fig. 1). In our computations, we use the following elongation schema. The last segment is considered as the growth cone with fixed length *d*. Changing the neurite length occurs due to the elongation of the next to the last segment. Its length is initially set to $l1$ which is the minimal segment length. If tubulin concentration is sufficient the length of the segment increases up to the value $d+l1$. After that the segment is divided on two parts with one segment of the length *d* and the other one next to the growth cone of the length $l1$.

## 3. Results

Let us now investigate the influence of the extracellular molecules on the growth dynamics. For illustration, first we consider two different distributions of growth factor sources (Fig. 2 A and B) and chose the direction of the neurite growth shown by arrows. Note, that model (1)-(4) did not account for the neurite navigation and the growth direction was kept fixed during the simulations.

Figure 3A illustrates the growing neurite dynamics in the profile shown in Fig. 2A. After some time of initial increase its length started to oscillate when the influence of the extracellular diffusive field became significant. The amplitude of the fluctuation decreased and when the neurite enters the region with a positive extracellular signal, it promoted the tubulin production and neurite growth is continued.

Surprising effect was found in the second case (the distribution is shown in Fig. 2B), where the length fluctuation converged to self-sustained limit cycle oscillations (Fig. 3B). In fact, the neurite became trapped within a non-uniform pattern of the extracellular field of signaling molecules. At the lapse of time the profile of the field tended to its equilibrium (Fig. 2B), but the neurite remained oscillating with certain amplitude and frequency (Fig. 3B).

The mechanism of the oscillatory dynamics can be explained by the interplay between promoting and depressing extracellular signaling. If the effective concentration is positive at the point of the growth cone $(x,y,t)$the feedback is positive and the tubulin production rate increases, that speeds up the elongation according to the Eqs. (1). Oppositely, when the effective concentration $S$ is negative the tubulin production rate decreases and the neurite slows down its growth. The most interesting case occurs when the growth cone concentration of tubulin reaches its threshold value, $C_n^{cr}=\beta/\alpha$. In this case the neurite starts its shortening, e.g. $dL/dt<0$, following the fourth equation of Eqs. (1).Next, if the length satisfies the condition:

$$L \leq d\,(n-1)+l1, \tag{5}$$

we assume that the length of the next to the last segment becomes less than minimal length, $l1$, and the $(n-1)$th and $(n-2)$th segments should be merged in one with concentration:

$$C_{n-1}(t) = \frac{C_{n-2} \cdot d + C_{n-1}(t-0) \cdot l}{d+l} \tag{6}$$

where $C_{n-1}(t-0)$ is the concentration in the last to the next segment before the mergence. So that the total neurite length decreases. It corresponds to the retraction phase of the length oscillations shown in Fig. 3.

Biological mechanisms underlying neurite pruning are still poorly understood. Like neurite elongation the process of retraction is based on cytoskeleton modifications. These transformations are regulated by signaling pathways that are activated from the extracellular environment or from cell-intrinsic programs[14]. In recent studies it has been shown that high concentration of the glutamate triggers axonal reduction[16]. Lysophosphatidic acid also induces neurite retraction via actin-dependent microtubule rearrangement[13]. Recent experimental data indicate that some signaling molecules are involved in the neurite retraction (for example Sema3F through neuropilin-2, plexin-A3, and plexin-A4 [15]).

Next, we illustrate the dependence of neurite elongation dynamics on different initial angles $\theta$ for the set of configurations of the growth factor sources shown in Fig. 2. As one may expect, for certain configuration the ratio between promoting and depressing influences depends on the growth direction and, hence, each angle would entail different neurite length dynamics. This is consistent with modern hypothesis that such inhomogeneity of the attractant concentration gradient (e.g. neurotrophic factor) determines the specificity of the cellular interactions. However, in the configuration presented in Fig. 2B neurites with any initial direction get locked at their certain length and start to oscillate as illustrated in Fig. 4B. Interestingly the oscillation frequency was angle-dependent and monotonically decreased with the increase of the neurite length. Note, that length fluctuation dynamics have been earlier reported in experimental and theoretical studies[27,31].

Finally, let us consider how the other two configurations shown in Fig 2 A and C provide a directional selectivity in the elongation dynamics. Figure 4A shows that depending on the growth angle the neurite may display either the outgrowth or the oscillations trapped by the profile of the extracellular field (Fig. 2C). In Figure 4C one can find that in addition to outgrowth and oscillations there are certain angle sector for which all neurites were completely suppressed.

## 4. Discussion

Neurite growth represents a key process in formation of morphological patterns in neural development. It was demonstrated in many experimental and theoretical models that the neurite length dynamics is driven mostly by intracellular signaling based on the production and transport of building proteins. In large scale computations such processes are often modeled by simple phenomenological functions [32,33]. Recent experiment shave suggested an important role of extracellular factors in modulation of the neurite development. Most of the studies typically discussed the extracellular guiding role in the navigation of the growth cone and influence on the probability of neurite branching. In addition to that our model has predicted that extracellular signaling can be also crucial in the growth dynamics. Supplying basic compartmental model by extracellular feedback affecting tubulin production rate we found that growing neurite can generate spontaneous oscillations defined by a non-uniform extracellular diffusive field. These extracellular molecules induce polymerization of building proteins in the growth cone and increase of cytoskeletal proteins synthesis. Neurotrophic factors also activate cell metabolism and accelerate transport of building proteins. On the contrary, increasing the concentration of agents activating protease-pathway and degradation of tubulin and actin can trigger axon retraction process.

Another interesting prediction by the model concerns the angle-dependent selectivity of the neurite elongation dynamics. In fact, the extracellular space is structured by the growth factor concentration on the areas preferable or non-preferable for the neurite outgrowth. It underlies the formation of certain neurite morphology in development together with the growth cone navigation and branching.

## 5. Conclusion

Neurite growth model based on the description of building protein production and supplied with extracellular feedback was proposed to explain neurite elongation and retraction processes. We showed that depending on the extracellular feedback and on neurite space orientation the proposed model generated different types of growth dynamics including monotonic outgrowth, oscillations locking the neurite in certain space region and complete inhibition of the growth. We also demonstrated that the elongation dynamics is angle-specific providing preferable or non-preferable directions of the neurite outgrowth.

## Acknowledgments

The research was partially supported by the Ministry of education and science of Russia (Project No.11.G34.31.0012), by the MCB Program of RussianAcademy of Sciences and by the Russian Foundation forBasic Research (Project Nos. 13-02-01223, 13-04-01871, 13-04-12067) and by the Russian President Grant No. MK-4602.2013.4. AYS and MVV appreciate Council for grants of the President of Russian Federation (СП-991.2012.4 and СП-411.2012.4) for financial support.

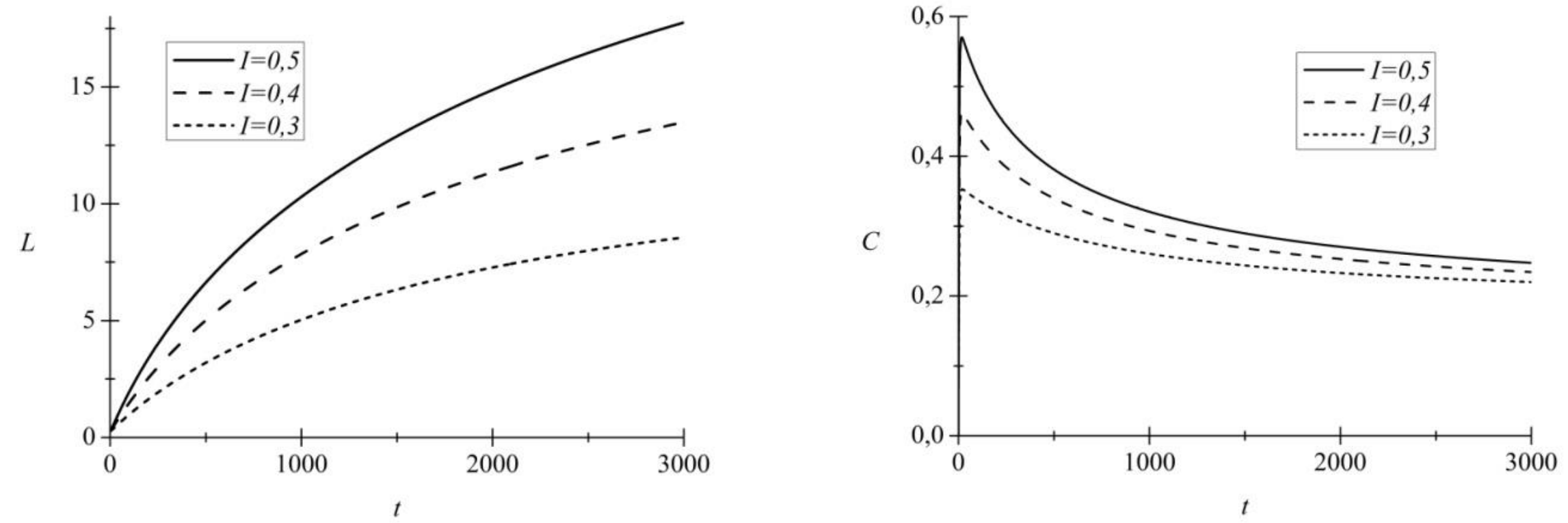

A B

Figure 1. Evolution of the neurite length (A) and tubulin concentration (B) in the growth cone for different values of tubulin production rate. Parameter values: $\gamma_0$=$\gamma$=0.4, $\alpha$=0.05, $\beta$=0.01, $D$=0.5, $l$1=0.1.

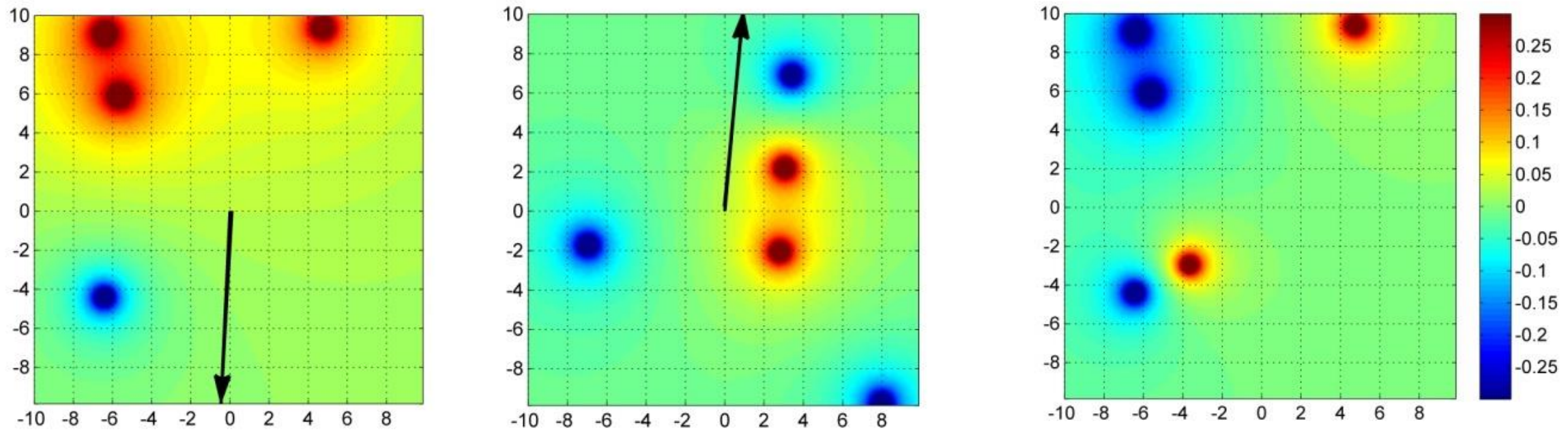

A B C

Figure 2. Configurations of extracellular signal sources used in simulations. The sources of growth promoting and depressing factors are shown by red and blue, respectively. The angles of neurite growth (unchanged in the current model) are shown by arrows. Color maps show equilibrium field of the growth factor formed for $t \to \infty$.

309

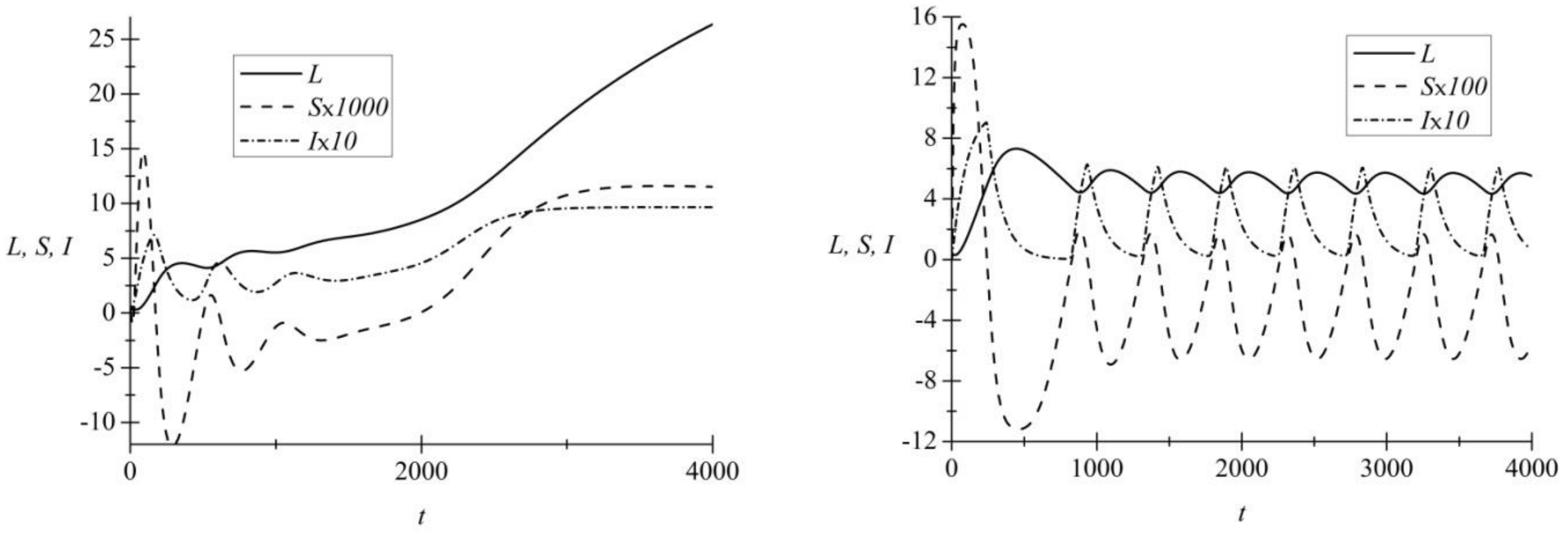

A B

Figure 3. Time evolution of neurite length, $L(t)$, effective concentration of growth factors, $S(x,y,t)$, in the growth cone and tubulin production rate, $I(t)$, in simulations of Eqs. (1)-(4), for sample distribution of extracellular signal sources (Fig. 2 A and B with the growth directions pointed by arrows) shown in panel A and B, respectively.Parameter values: $d$=0.2, $\gamma_0$=$\gamma$=0.4, α=0.05, $\beta$=0.01, $D$=0.5, $\Delta t$=0.01, $l1$=0.1, $I^*$=0.5, $q$=1, $D_1$=0.5, $\delta$=0.01, $A$=0.01, $B$=0, $C$=0.01.

310

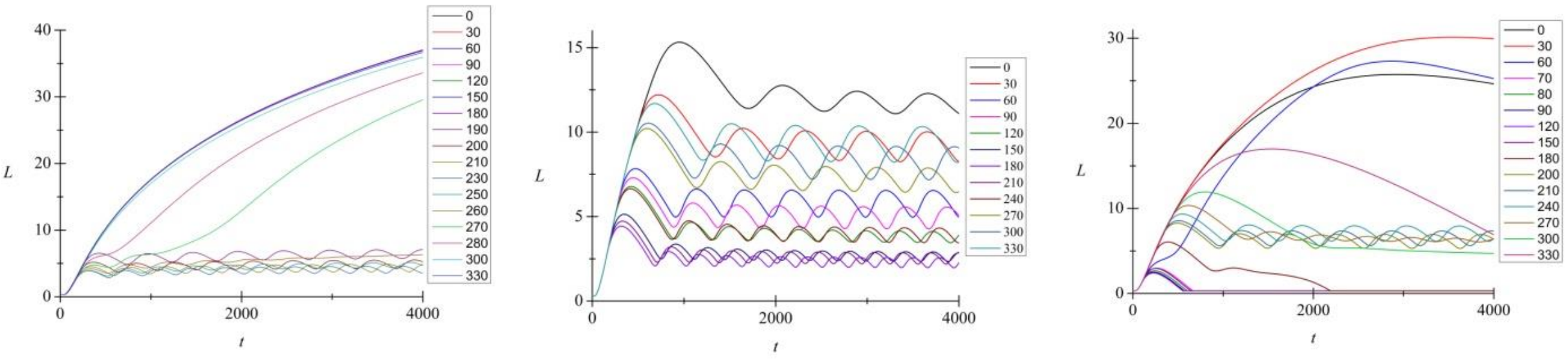

A B C

Figure 4. Neurite elongation dynamics for different growth angles (counterclockwise) in growth factor sources configurations shown in Fig. 2 A, B, C (on A, B and C respectively). Parameter values: $d$=0.2, $\gamma_0$=$\gamma$=0.4, $\alpha$=0.05, $\beta$=0.01, $D$=0.5, $\Delta t$=0.01, $l1$=0.1, $I^*$=0.5, $q$=1, $D_1$=0.5,$\delta$=0.01, $A$=0.01, $B$=0, $C$=0.01.

311